\documentclass{iopjournal}

\usepackage{siunitx}
\usepackage{subcaption}
\usepackage{booktabs}
\usepackage{float}
\usepackage{indentfirst}
\usepackage{amsmath}      
\usepackage{amssymb}     
\usepackage{mathtools}
\begin{document}

\articletype{Article type}

\title{Patch-MLP-Based Predictive Control: Simulation of Upstream Pointing Stabilization for PHELIX Laser System}

\author{
Jiaying Wang\textsuperscript{1*}\orcid{0009-0003-2576-1623}, 
Jonas Benjamin Ohland\textsuperscript{2}\orcid{0000-0002-1328-2188}, 
Yen-Yu Chang\textsuperscript{3}\orcid{0000-0003-0548-3999}, 
Vedhas Pandit\textsuperscript{1}\orcid{0000-0002-1983-8140}, 
Stefan Bock\textsuperscript{1}\orcid{0000-0002-1919-8585}, 
Andrew-Hiroaki Okukura\textsuperscript{4,5,6}\orcid{0000-0002-4722-5894}, 
Udo Eisenbarth\textsuperscript{2}\orcid{0009-0001-0168-5932}, 
Arie Irman\textsuperscript{1}\orcid{0000-0002-4626-0049},
Michael Bussmann\textsuperscript{1}\orcid{0000-0002-8258-3881}, 
Ulrich Schramm\textsuperscript{1,7}\orcid{0000-0003-0390-7671}, 
Jeffrey Kelling\textsuperscript{1,8}\orcid{0000-0003-1761-2591}
}

\affil{$^1$Helmholtz-Zentrum Dresden-Rossendorf, Dresden, Germany}
\affil{$^2$GSI Helmholtzzentrum für Schwerionenforschung, Germany}
\affil{$^3$Amplitude laser group - Dresden operations, Germany}
\affil{$^4$Extreme Light Infrastructure - Nuclear Physics, National Institute for Physics and Nuclear Engineering, Ilfov, Romania }
\affil{$^5$University of Bucharest, Ilfov, Romania }
\affil{$^6$ Engineering and Applications of Lasers and Accelerators Doctoral School (SDIALA), National University of Science and Technology Politehnica of Bucharest, Bucharest RO-060042, Romania }
\affil{$^7$Technische Universität Dresden, Dresden, Germany}
\affil{$^8$Chemnitz University of Technology, Chemnitz, Germany }

\affil{$^*$Author to whom any correspondence should be addressed.}

\email{j.wang@hzdr.de}


\begin{abstract}
High-energy laser facilities such as PHELIX at GSI require excellent beam pointing stability for reproducability and relative independence for future experiments. Beam pointing stability has been traditionally achieved using simple proportional–integral–derivative (PID) control which removes the problem of slow drift, but is limited because of the time delay in knowing the diagnosis and the inertia in the mechanical system associated with mirrors. In this work, we introduce a predictive control strategy where the forecasting of beam pointing errors is performed by a patch-based multilayer perceptron (Patch-MLP) designed to capture local temporal patterns for more robust short-term jitter prediction. The subsequent conversion of these predicted errors into correction signals is handled by a PID controller. The neural network has been trained on diagnostic time-series data to predict beam pointing error. Using the hybrid predictive–feedback controller compensates for system delays. Simulations with a correction mirror placed upstream of the PHELIX pre-amplifier bridge confirm that the predictive control scheme reduces residual jitter compared to conventional PID control. In our simulations, over a 10-hour dataset the controller maintained stable performance without drift, while standard pointing metrics showed consistent improvements of the order of $10\%\text{--}20\%$. The predictive controller operates without drift, and therefore may improve reproducibility and operational efficiency in high energy, low repetition rate laser experiment conditions.
\end{abstract}

\section{Introduction}

High-power laser facilities have become essential for advancing research in high-energy-density physics, plasma acceleration, and extreme states of matter~\cite{major2024phelix,Leemans2010}. Systems such as PHELIX at GSI and BELLA at LBNL demand exceptional beam quality and stability to ensure reproducibility in applications ranging from laser-plasma acceleration to warm dense matter studies~\cite{major2024phelix,berkeley2017}. Within such facilities, even small mechanical vibrations or thermal drifts can be transmitted through structural components of the beamline, making laser pointing highly sensitive to environmental disturbances. This directly impacts experiments with very small targets, such as wire irradiation~\cite{Akstaller2021}, where micrometer-scale displacements already reduce reproducibility. Furthermore, pointing instabilities degrade the performance of beam correction systems: once steep focusing optics are introduced—such as the final focusing element or the Petawatt Target Area Sensor (PTAS) at PHELIX which includes an off-axis parabolic telescope~\cite{Ohland2022}— pointing errors are effectively converted into higher-order aberrations, including astigmatism. As a result, beam jitter constitutes a fundamental limitation to attainable beam quality and downstream experimental reliability~\cite{major2024phelix,berger2023,Jensen2025}.

\begin{figure}[t]
  \centering
  \includegraphics[width=1\textwidth]{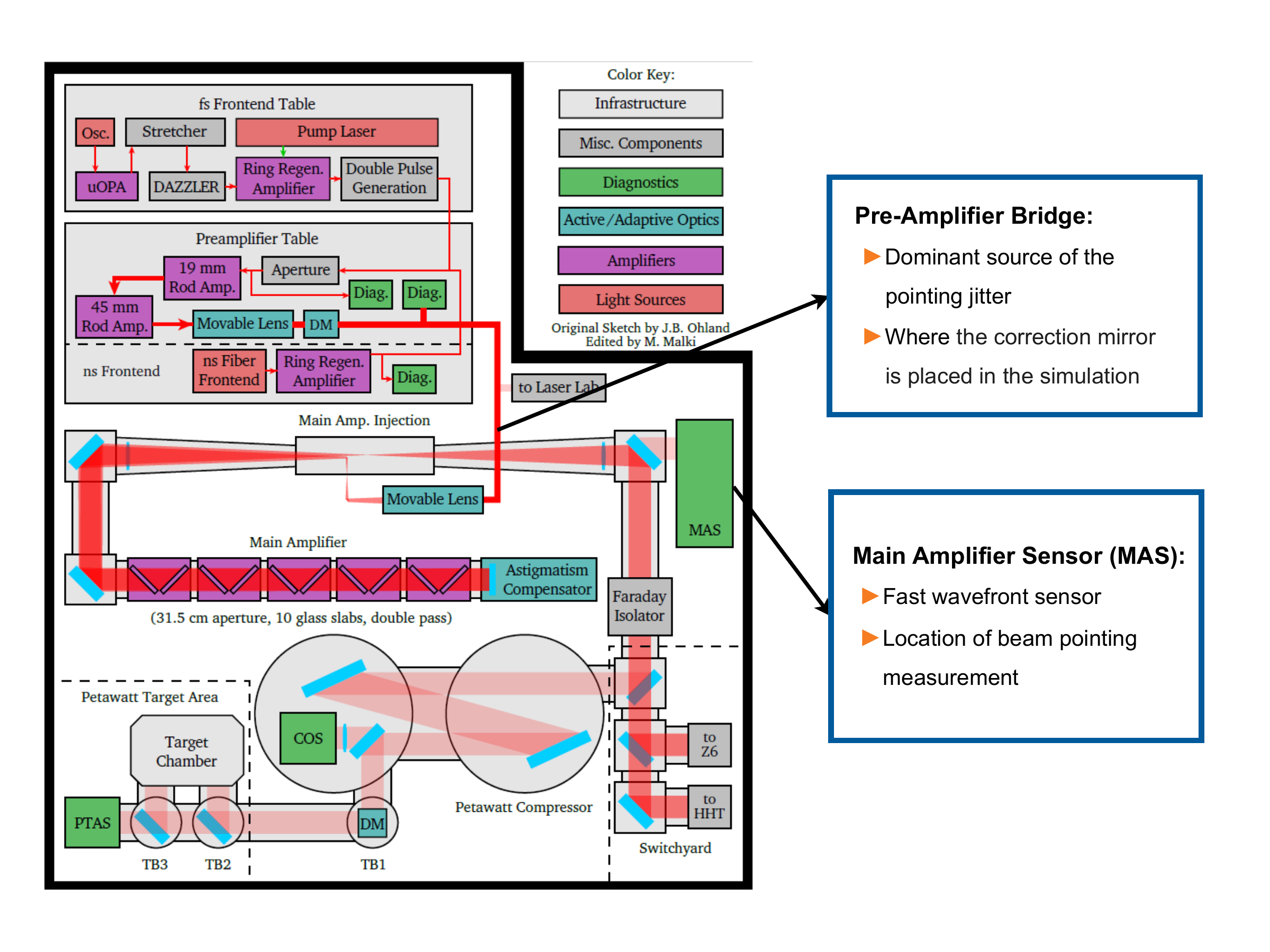}
  \caption{Simplified layout of the PHELIX~\cite{major2024phelix} laser system highlighting the dominant source of pointing jitter (pre-amplifier bridge) and the location of the pointing measurement (Main Amplifier Sensor, MAS). These positions provide the physical background for the simulation study presented in this work.}
  \label{fig:phelix_layout}
\end{figure}

Mitigation strategies, such as active pointing stabilization with fast steering mirrors, are theoretically most effective when applied as close as possible to the source of jitter, since compensation in different image planes inevitably introduces near field shifts. Nevertheless, deliberate pre-compensation earlier in the beamline is often practiced, as the inertia of larger steering mirrors limits their mechanical response and makes fast correction impractical with conventional controllers~\cite{major2024phelix}. Conventional stabilization schemes rely on proportional–integral–derivative (PID) control loops, which are well established in engineering practice and effective at suppressing slow drifts and low-frequency disturbances~\cite{franklin2002,genoud20}. However, PID control inherently acts reactively on past or present errors, and its bandwidth is fundamentally limited by the inertia of large correction optics~\cite{berger2023,oppenheim2010}. As a result, high-frequency disturbances in the tens to hundreds of Hz range remain only partially compensated~\cite{isono2021,wu2020}. This leads to residual pointing instabilities that exceed the stringent tolerances required for applications such as laser-plasma accelerators or ion-beam coupling experiments~\cite{maier2020,gonsalves2015}.

In the case of the PHELIX laser at GSI, the pre-amplifier bridge (see Fig.~\ref{fig:phelix_layout}) is the dominant source of pointing jitter due to vibrations in the ground. Pointing is routinely diagnosed downstream at the Main Amplifier Sensor (MAS), which includes a fast wavefront sensor located after the main amplifier chain. These two locations define the physical background for the present simulation study: while the MAS provides the experimental reference point for pointing stability, the pre-amplifier bridge represents the most relevant source of perturbations where an active correction mirror could in principle be introduced in the future. 

In this work, we introduce a new machine-learning-enhanced control strategy that integrates predictive modeling with PID feedback. By training a MLP on time sequences of diagnostic signals, the controller is able to forecast future pointing errors and provide the PID loop with a preemptive correction signal. This predictive capability transforms the inherently lagging behavior of conventional PID into a forward-looking control scheme, effectively extending the usable bandwidth beyond hardware limitations. As we will demonstrate through numerical simulations and experimental data analysis, the MLP-driven PID maintains stable performance over a 10-hour dataset, achieving reductions of $17\%\text{--}20\%$ in root-mean-square (RMS) pointing error, $10\%\text{--}13\%$ in peak-to-valley (P2V), and $18\%\text{--}21\%$ in full width at half maximum (FWHM). These results directly characterize the long-term stability of the method and underline its potential for improving reproducibility in high-power, low-repetition-rate laser systems.

\section{Methodology}
\subsection{Workflow of the simulation }

\indent PID control continues to be one of the most extensively implemented strategies in precision engineering and laser stabilization systems. The discrete-time PID controller can be expressed as follows:

\begin{equation}
u[k] = K_p\, e[k]
+ K_i \sum_{i=0}^{k} e[i]
+ K_d \frac{e[k] - e[k-1]}{T_s}
\label{eq:pid_position}
\end{equation}
where these three parameters ($K_p,K_i,K_d$) can be adjusted. 

Within high-energy laser systems, PID controllers are routinely used to drive corrective tip–tilt mirrors to reduce slow drifts of the laser beam position that originate from thermal gradients, air turbulence, or gradual mechanical deformation. Such controllers are very efficient at attenuating long-term pointing drifts and very low-frequency fluctuations ($<$ 1 Hz)~\cite{berger2023,BeamStab2025,Amodio2025}, which helps maintain stable beam delivery to the target. However, their performance is inherently restricted by the inertia and limited bandwidth of the mirror, especially if larger and thus heavier mirrors are used (e.g. 4 inch diameter), which  limits their ability to suppress higher-frequency jitter components ($>$ 10–20 Hz)~\cite{berger2023,Amodio2025}.

\begin{figure}[t]
 \centering
        \includegraphics[width=0.9\textwidth]{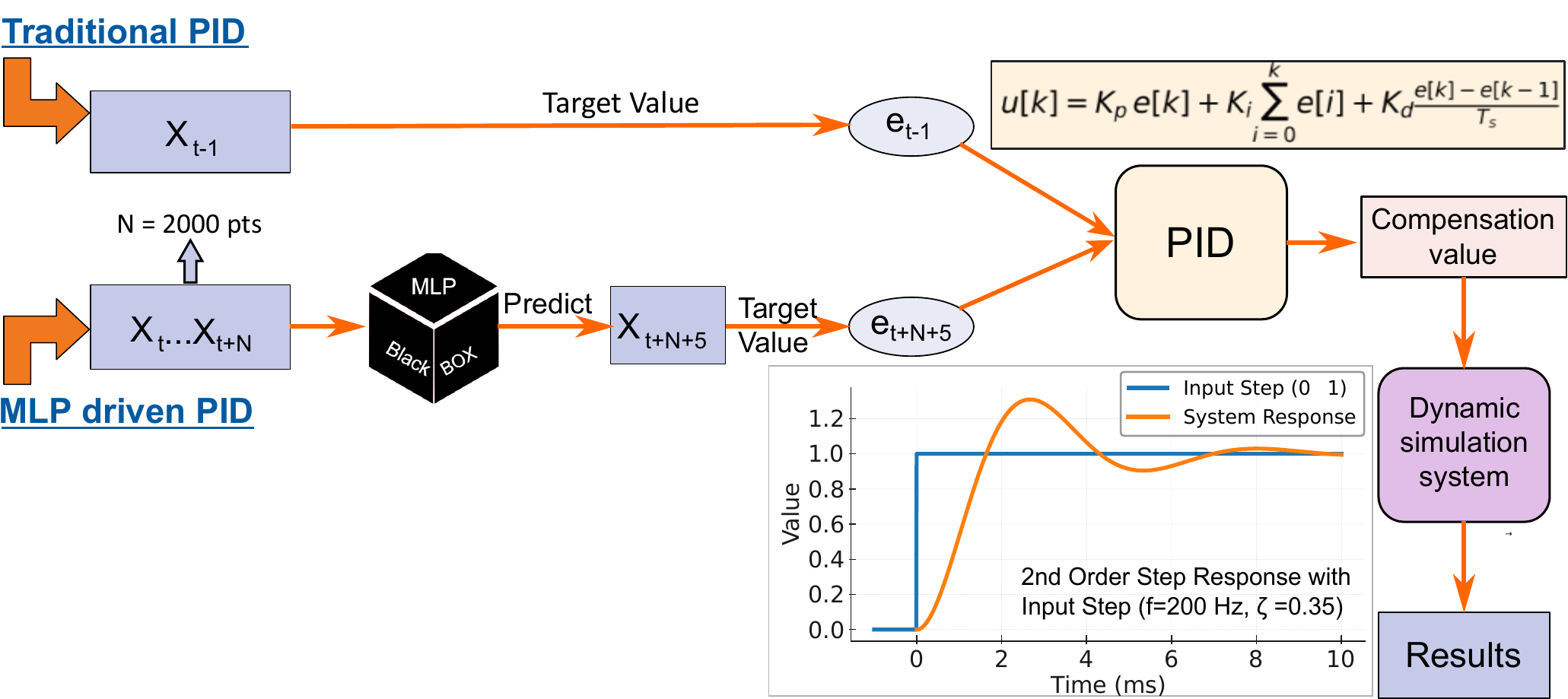}
 \caption{Workflow of the Hybrid predictive–feedback MLP-Driven PID and Traditional PID Control. The inset shows the step-response of the steering mirror model used in our simulation.}
 \label{fig:WORKFLOW}
\end{figure}

As illustrated in Fig. \ref{fig:WORKFLOW}, we present a hybrid predictive–feedback control scheme that combines an MLP time-series forecaster with a traditional PID controller to realize predictive beam pointing stabilization. The predicted positions are then converted into commands for a tip–tilt mirror at the pre-Amplifier bridge.

As shown in Fig. \ref{fig:WORKFLOW}, we use a second-order damped transfer function to represent the piezo-actuated tip-tilt mirror. The dynamics of the piezo-actuated tip-tilt mirror are approximated by a second-order mass–spring–damper system, as the relatively large mirror (e.g. 4 inch) mass dominates the response. This results in a single resonance mode that captures the essential dynamics, while higher-order modes of the piezo actuator are considered negligible within the operating frequency range. This lets us describe how its mechanical behavior affects the correction performance during simulation. The transfer function in Laplace space given by $G(s) = \omega_n^2/(s^2 + 2 \zeta \omega_n s + \omega_n^2$). $\omega_n = 2 \pi f$ denotes the natural angular frequency of the mirror, where $f = 200~\mathrm{Hz}$, and $\zeta = 0.35$ is the damping ratio. The parameter values were chosen as a pragmatic approximation for a large 4 inch tip/tilt mirror, based on extrapolations from the PI S-340 datasheet towards heavier mirrors, sufficient for a demonstration model~\cite{PI_S340}.

The simulated compensation value is superimposed on the measured laser beam center position to obtain the simulated corrected trajectory. An important point here is that the traditional PID approach introduces a system delay of 0.6 ms, chosen in analogy to the latency reported for the Apollon ARTAO system~\cite{Ohland2025}. In contrast, our new method applies the predicted correction in advance, making the response delay-free.

\subsection{Patch MLP}
Machine learning (ML) is a time-series forecasting method to predict the future behavior of a dynamical system. It has been applied in diverse fields such as finance, climate modeling, and industrial process control. Recently, it has been demonstrated that deep learning models can accurately predict the center position of a laser beam. These studies~\cite{Amodio2025, Berger2025} imply that the prediction time only needs to cover the model computation time and the optical system delay (dominated by the latency of the driver)~\cite{Ohland2025}. Therefore, aiming for short prediction intervals, we employed a relatively lightweight architecture—an MLP—which provides sufficient predictive capability while maintaining low computational time. Transformer-based architectures are avoided because they are computationally more expensive, and recent studies show that their permutation-invariant self-attention fails to reliably capture temporal order in physical time-series, while simpler models often achieve equal or better short-horizon performance~\cite{zeng2023}.

To further improve accuracy, a Patch MLP~\cite{tang2025patchmlp} was adopted: the input sequence is divided into fixed-length patches, and by adjusting the patch size, the model can be tuned to better capture the frequency components most relevant for beam jitter prediction.

\begin{figure}[t]
    \centering
    \includegraphics[width=0.8\linewidth]{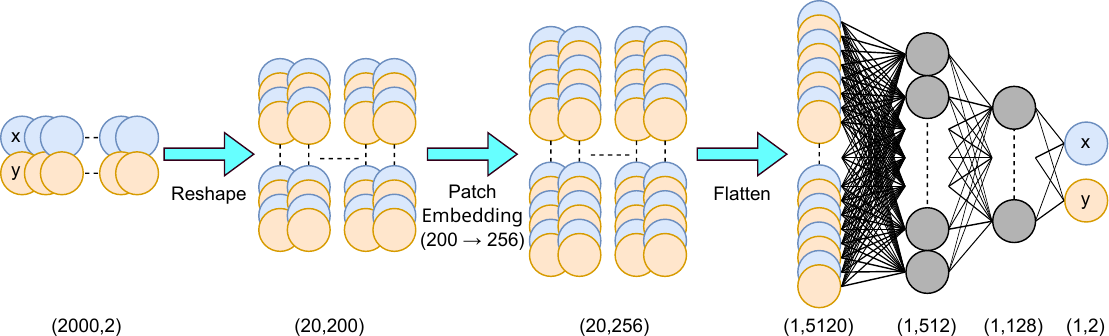}
    \caption{Patch-based MLP Architecture for Two-Axis Beam Pointing Prediction.}
    \label{fig:patch based MLP}
\end{figure}

\paragraph{Input and Preprocessing}

As shown in Fig. \ref{fig:patch based MLP}, We use a sliding window of length $N=2000$ composed of two channels (horizontal $x_i$ and vertical $y_i$ pointing), arranged as
\begin{equation}
\mathbf{X} =
\begin{bmatrix}
x_1 & x_2 & \cdots & x_{2000} \\
y_1 & y_2 & \cdots & y_{2000}
\end{bmatrix}
\in \mathbb{R}^{2 \times N}.
\label{eq:input}
\end{equation}

 Before being fed into the network, the input is normalized to $[0,1]$ using min--max scaling. 

\paragraph{Patch Partitioning and Embedding}
The sequence is divided into $M = N/P$ non-overlapping patches of length $2P = 200$.
Each patch is flattened into a vector:
\begin{equation}
\mathbf{p}_i=\mathrm{vec}\!\big(\mathbf{X}[:,(i-1)P+1:iP]\big)\in\mathbb{R}^{2P},
\qquad i=1,\dots,M.
\label{eq:patch}
\end{equation}
A linear projection maps every patch to a $d$-dimensional embedding ($d=256$):
\begin{equation}
\mathbf{z}^{(0)}_i=\mathbf{W}_e\,\mathbf{p}_i+\mathbf{b}_e \in \mathbb{R}^{d},
\qquad i=1,\dots,M .
\label{eq:embed}
\end{equation}

\paragraph{Feature Concatenation and MLP}

Patch embeddings are concatenated into a global representation:
\begin{equation}
\mathbf{h}^{(0)}=\mathrm{concat}\!\big(\mathbf{z}^{(0)}_1,\dots,\mathbf{z}^{(0)}_M\big)
\in\mathbb{R}^{M\cdot d}.
\label{eq:concat}
\end{equation}
The vector $\mathbf{h}^{(0)}$ is processed by a MLP with ReLU activations:
\begin{align}
\mathbf{h}^{(1)} &= \sigma\!\big(\mathbf{W}_1\mathbf{h}^{(0)}+\mathbf{b}_1\big),
& \mathbf{h}^{(1)}\in\mathbb{R}^{512}, \label{eq:mlp1}\\
\mathbf{h}^{(2)} &= \sigma\!\big(\mathbf{W}_2\mathbf{h}^{(1)}+\mathbf{b}_2\big),
& \mathbf{h}^{(2)}\in\mathbb{R}^{128}, \label{eq:mlp2}\\
\hat{\mathbf{y}} &= \mathbf{W}_3\mathbf{h}^{(2)}+\mathbf{b}_3,
& \hat{\mathbf{y}}\in\mathbb{R}^{2}. \label{eq:out}
\end{align}
The final output $\hat{\mathbf{y}}=[\,\hat{x},\,\hat{y}\,]^{\top}$ represents the predicted horizontal and vertical beam positions at a future time step.

\paragraph{Loss function}
The model is trained to minimize the mean–squared error (MSE) between the normalized prediction and the ground-truth coordinates $(x,y)$ at an prediction offset of $\Delta = 5$ points. For a mini-batch $\{(\mathbf{X}^{(b)}, \mathbf{y}^{(b)})\}_{b=1}^B$,
with $\hat{\mathbf{y}}^{(b)}\in\mathbb{R}^2$ and $\mathbf{y}^{(b)}\in\mathbb{R}^2$ (both after min--max scaling), the loss is:
\[
\mathcal{L}_{\mathrm{MSE}}
= \frac{1}{B}\sum_{b=1}^{B} \left\|\hat{\mathbf{y}}^{(b)}-\mathbf{y}^{(b)}\right\|_2^2 
\]
where $B$ is the mini-batch size, and $\mathbf{y}^{(b)} = (x^{(b)}, y^{(b)})$ contains the ground-truth future coordinates and $\hat{\mathbf{y}}^{(b)}$ denotes the model prediction.

The two axes are equally weighted. During evaluation, predictions are inverse-scaled with the training min–max parameters to restore their physical units before computing metrics. 

\paragraph{Sliding-Window Prediction}
During inference, the input window is advanced by one sample at each step, producing a continuous stream of predictions:
\begin{equation}
f_{\theta}:\ \mathbb{R}^{2\times 2000}\rightarrow\mathbb{R}^{2},\qquad
\hat{\mathbf{y}}_{t+\Delta}=f_{\theta}\!\big(\mathbf{X}_{t:t+N-1}\big),
\label{eq:mapping}
\end{equation}
where $\theta$ denotes all trainable parameters. We use time-series data acquired by the PHELIX MAS \cite{major2024phelix} with a sampling interval of $T_s=0.244~\mathrm{ms}$ (sampling period). 
To account for both the per-step inference time of the model and the system latency, we set the prediction offset to $\Delta=5$ points, i.e., a horizon of $\Delta T_s \approx 1.22~\mathrm{ms}$, chosen according to $\Delta > \left\lceil (T_{\text{model}} + T_{\text{sys}})/T_s \right\rceil$.
Predictions at offset $\Delta$ are then used for compensation.

\section{Simulation Results and Discussion}
Before presenting the results, we briefly summarize the training procedure. The Patch-MLP forecaster, with about $2\times10^6$ trainable parameters, was trained on one minute of time-series data sampled at $T_s=0.244$ ms (yielding $\sim 2.4\times10^5$ samples). The first $80\%$ of this sequence was used for training and the remaining $20\%$ for validation, ensuring a temporally consistent split without information leakage. The model was trained for 50 epochs, which took about three minutes  of wall-clock time on two NVIDIA V100 GPUs.
\subsection{Comparison of simulation results after compensation}

\begin{figure}[t]
    \centering
    \includegraphics[width=0.60\linewidth]{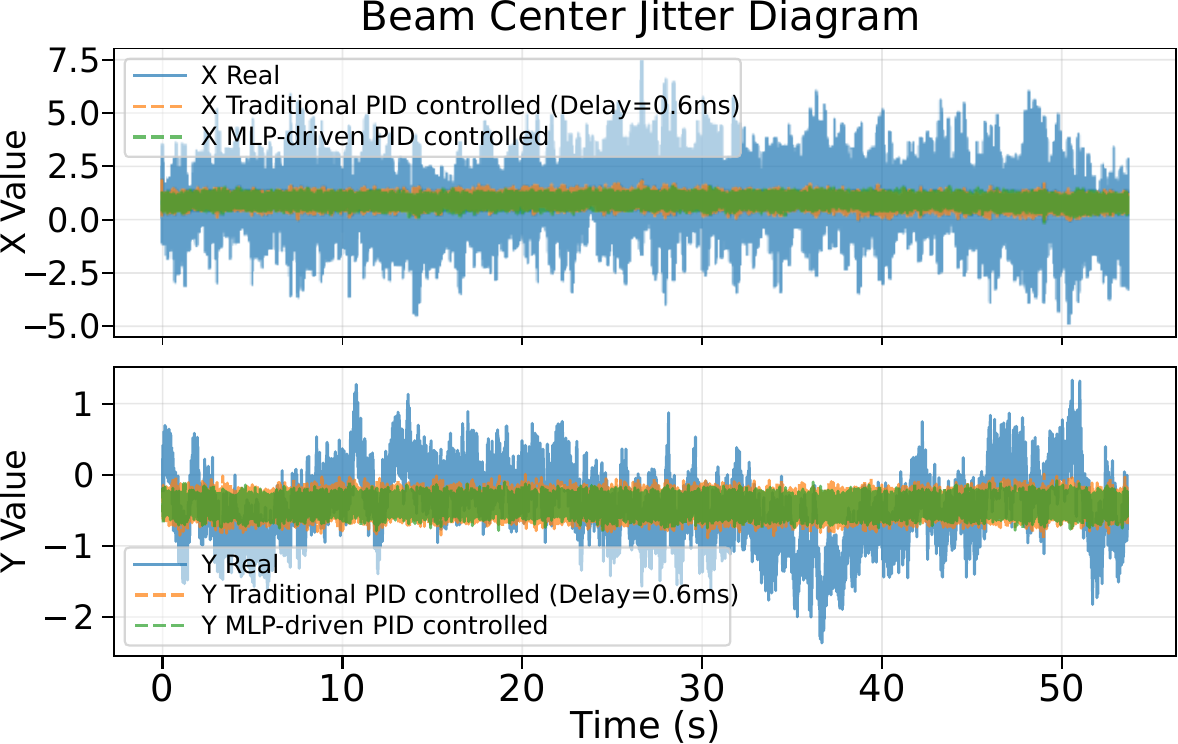}
    \caption{One-minute beam-center trajectories (X and Y): real signal vs. simulated traditional PID (0.6\,ms delay) vs. simulated MLP-driven PID.}
    \label{fig:timeseries_1min}
\end{figure}

Simulation results, presented in Fig.~\ref{fig:timeseries_1min}. The blue trace shows the uncontrolled signal; the orange trace is the output under a traditional PID controller with a $0.6\,\mathrm{ms}$  delay time; the green trace is the proposed MLP-driven PID that uses a prediction offset of $\Delta=5$ points ($\approx1.22\,\mathrm{ms}$) to apply zero-delay corrections. Both controllers markedly reduce the amplitude envelope compared with the real signal. And the proposed MLP-driven PID controller has better performance throughout the entire one-minute data.

\begin{figure}[t]
    \centering
    \begin{subfigure}{0.55\linewidth}
        \centering
        \includegraphics[width=\linewidth]{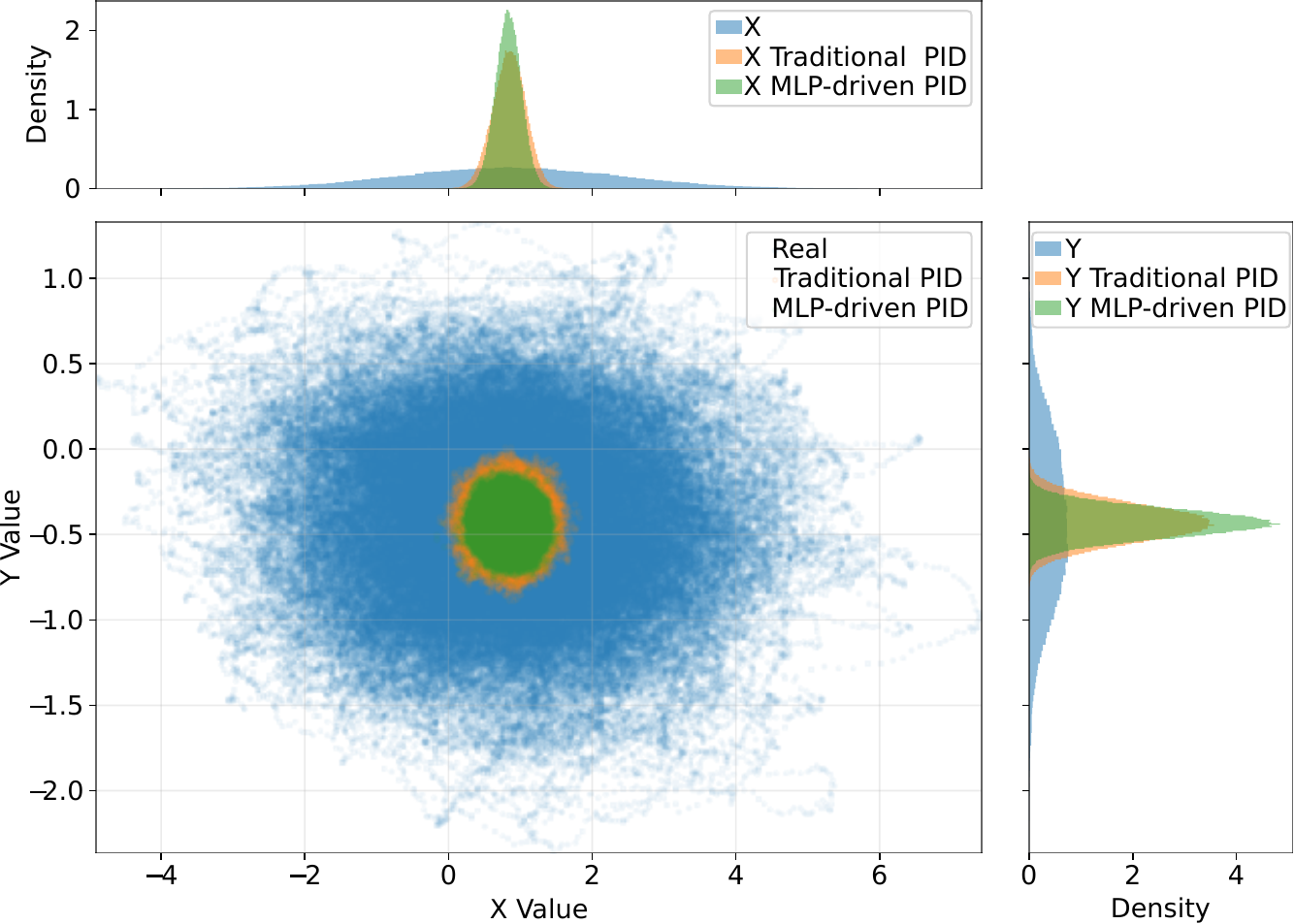}
        \caption{Beam-center distribution with marginal densities: uncontrolled (blue), traditional PID (orange), and MLP-driven PID (green).}
        \label{fig:dist_marginals}
    \end{subfigure}
    \hfill
    \begin{subfigure}{0.42\linewidth}
        \centering
        \includegraphics[width=\linewidth]{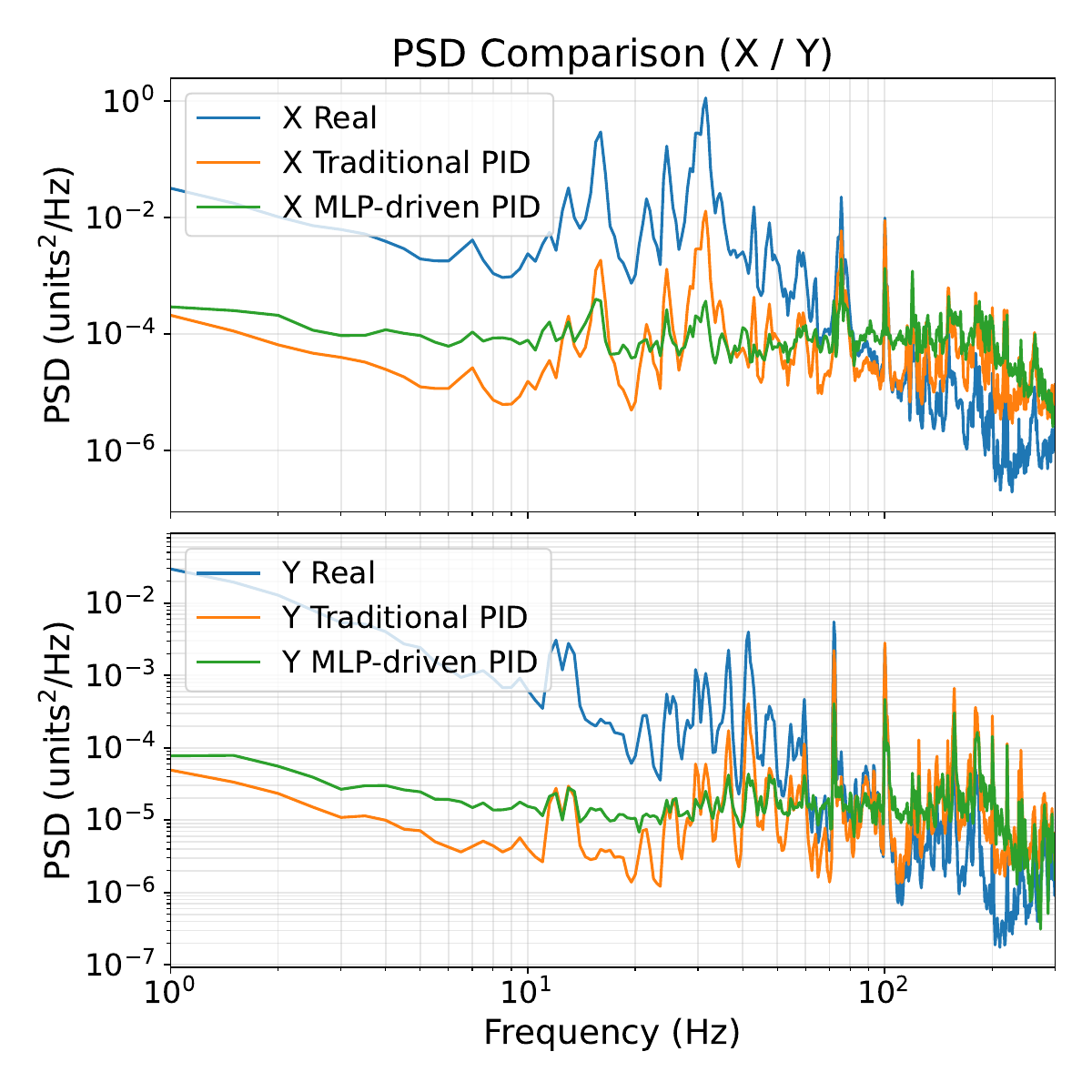}
        \caption{Power spectral density of beam jitter (X/Y, 0–300\,Hz): comparison of real, traditional PID, and MLP-driven PID.}
        \label{fig:psd}
    \end{subfigure}
    \caption{Comparison of simulated beam stabilization: (a) beam-center distribution and (b) power spectral density.}
    \label{fig:results_combined}
\end{figure}

Fig.~\ref{fig:dist_marginals} shows the improvement in laser pointing stability brought about by the new method on the same one-minute dataset as used in Fig.~\ref{fig:timeseries_1min}, recorded two hours and 15 minutes after the training data. Under the control of the new method, the distribution of the laser beam center point is more concentrated in both the x-axis and the y-axis compared to the traditional PID method. 

The MLP-driven PID (green) achieves a clear advantage below $60$ Hz, not through a uniform reduction over the entire band, but through a pronounced suppression of the dominant residual peaks that remain under conventional PID control on both axes (Fig.~\ref{fig:psd}). The improvement arises from the prediction offset ($\Delta=5$ samples $\approx 1.22$ ms), which compensates for loop and mirror delays and reduces the phase lag that constrains a feedback-only PID in this frequency band. Consequently, while the conventional loop cannot fully attenuate the sub-$60$ Hz content, the predictive method achieves markedly lower residual amplitudes. A comparison of cumulative RMS of feedback-only PID and MLP-driven PID can be found in the supplementary data.

In summary, the proposed MLP-driven PID consistently outperforms the traditional PID: the beam-center distribution is the tightest, the time traces show the smallest residual band without drift, and the PSD shows a uniformly lower broadband floor—most notably below 60\,Hz, where the traditional PID leaves significant residuals—without inducing oscillations, indicating compatibility with the mirror dynamics. Motivated by these short-term gains, we next examine long-term stability over extended runs.

\subsection{Model stability}

For the long-term simulation, a single model was trained on one 1\,min MAS segment and then kept fixed. The model was applied to all subsequent data collected over \(\sim 10\)\,h. Acquisition was intermittent rather than strictly periodic: most intervals were 5\,min, with occasional 10\,min gaps. In total, 110 valid 1\,min segments were obtained and used for evaluation. For each segment, predictions were generated and converted into simulated corrections following the procedure described above. The one-minute training segment and the 10\,h test dataset are fully disjoint, with the latter taken after the training data on the same day. No test-set information was used for model training, and no hyperparameter tuning was performed on the test data. All hyperparameters were fixed before evaluation, and the trained model was applied to the 10\,h dataset without further adjustment.

\begin{figure}[t]
    \centering
    \includegraphics[width=0.85\linewidth]{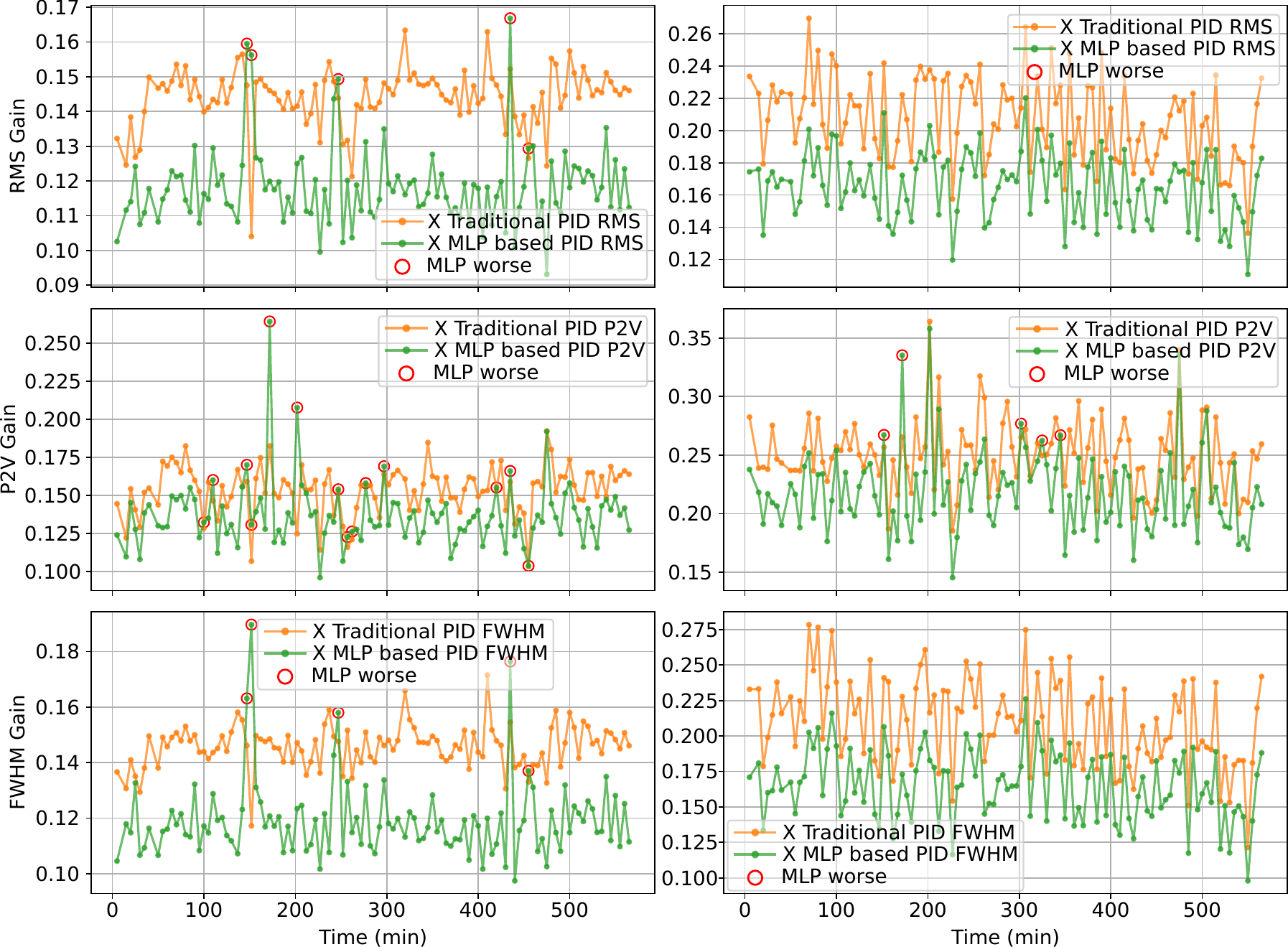}
    \caption{Average percentage reduction of residuals (MLP-driven PID vs.\ traditional PID) over the 10\,h simulated test. Positive values indicate lower residuals with the MLP-driven controller.}
    \label{fig:gain}
\end{figure}

As illustrated in Fig.~\ref{fig:gain}, we quantify performance using three dimensionless gain metrics -- RMS, P2V, and FWHM each defined as the ratio of the compensated residual to the original value, with values $< 1$ indicating improvement.

\begin{align}
G_{\mathrm{RMS}}^{(a)}   &= \frac{\mathrm{RMS}^{(a)}_{\mathrm{comp}}}{\mathrm{RMS}^{(a)}_{\mathrm{orig}}},\\
G_{\mathrm{P2V}}^{(a)}   &= \frac{\mathrm{P2V}^{(a)}_{\mathrm{comp}}}{\mathrm{P2V}^{(a)}_{\mathrm{orig}}},\\
G_{\mathrm{FWHM}}^{(a)}  &= \frac{\mathrm{FWHM}^{(a)}_{\mathrm{comp}}}{\mathrm{FWHM}^{(a)}_{\mathrm{orig}}},\\
\mathrm{Reduction}\,(\%) &= 
\frac{G_{\mathrm{Traditional\, PID}} - G_{\mathrm{MLP\text{-}driven\, PID}}}
     {G_{\mathrm{Traditional\, PID}}} \times 100\%.
\label{eq:reduction}
\end{align}

\paragraph{Result assessment}
Across the 110 one–minute segments (\(\sim10\,\)h), the MLP-driven PID consistently outperforms the traditional PID. The per-segment gain curves (Fig. \ref{fig:gain}) show the MLP-driven PID traces lying mostly below the traditional PID ones (smaller is better) with no upward drift, indicating stable performance without degradation over time. Averaged over all segments, the percentage reductions (MLP-driven PID vs.\ Traditional PID) are:

\begin{table}[t]
  \centering
  \begin{tabular}{lccc}
    \hline
    & RMS & P2V & FWHM \\
    \hline
    X direction & 17.2\% & 10.8\% & 18.2\% \\
    Y direction & 20.1\% & 12.5\% & 21.4\% \\
    \hline
  \end{tabular}
  \caption{Average relative improvement of the jitter reduction of the MLP-driven PID compared to the traditional PID, calculated by Eq.~\ref{eq:reduction}.}
\end{table}

\paragraph{Failure cases}
Occasional segments where the MLP-driven PID is worse (circled) are rare and typically coincide with outlier excursions that affect P2V. In a few minute-long segments we observe a simultaneous degradation of all three metrics (RMS, P2V, and FWHM), indicating a short-term failure of the predictive method.

\begin{figure}[t]
    \centering
    \includegraphics[width=0.7\linewidth]{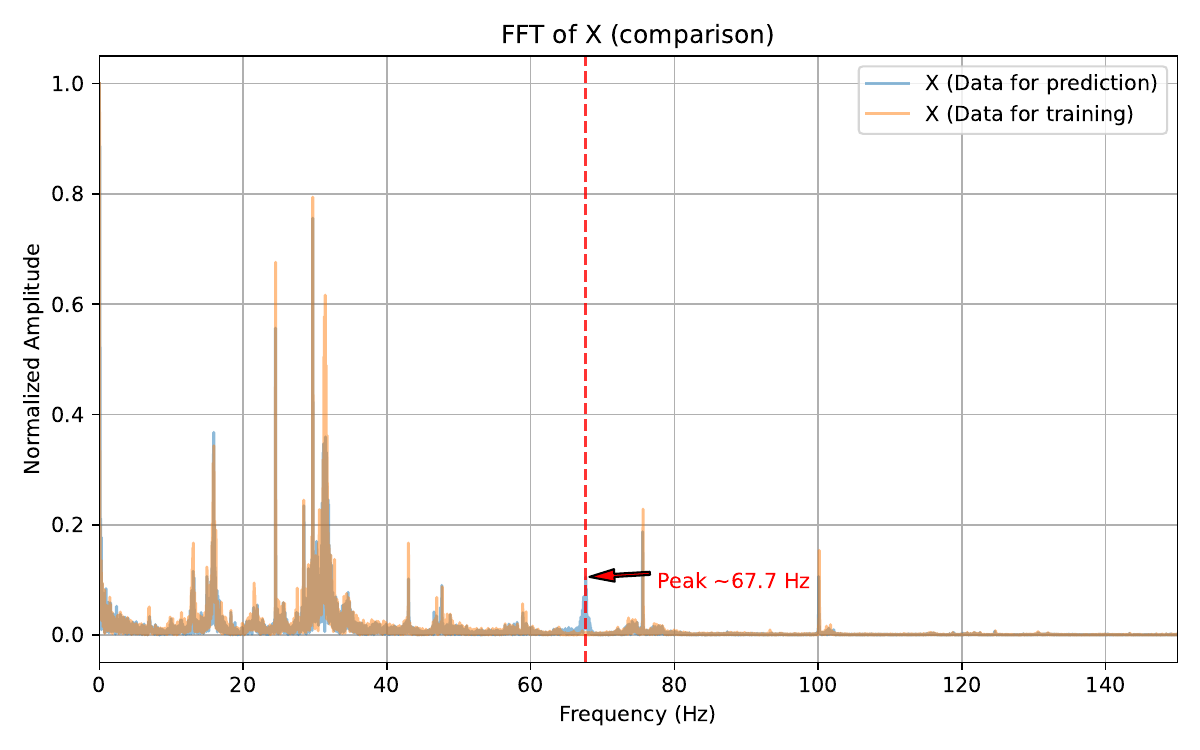}
    \caption{Fourier comparison of failure segments and training data.}
    \label{fig:worse}
\end{figure}

Fig.~\ref{fig:worse} compares the Fast Fourier Transform (FFT) spectra of a one-minute segment of x-axis data with the training set. A distinct peak emerges around 67.7 Hz, which is absent in both the training data and the well-predicted segments but consistently present in all minute-long intervals with degraded predictions. We therefore attribute this previously unseen spectral component as the likely cause of the model’s reduced prediction performance. We hypothesize that long-range motion of a gantry present near the laser system or other new external vibration sources introduced additional vibration modes, leading to a distribution shift and reduced prediction accuracy. This interpretation requires confirmation in dedicated subsequent experiments.

Overall, the predictive controller yields lower variance and a narrower distribution of the beam center on both axes, with no signs of drift across \SI{10}{h}. Because the model can be retrained within only a few minutes using approximately one minute of newly acquired data, it would also be straightforward to update the model should longer-term drift occur. In future work, the prediction accuracy could be further improved by distinguishing different classes of environmental vibration sources based on downstream experimental requirements, allowing the model to focus on the most relevant disturbance components.

\section{Conclusion}
This work illustrates the use of predictive control with a Patch-MLP forecaster to expand available PID-based laser pointing stabilization bandwidth. This method operates by predicting which future errors are most likely and applying a correction without time delays to the laser output. The forecaster routinely outperforms standard PID for RMS, P2V, and FWHM residuals during both short-term and long-term reporting. We show that the results are steady and repeatable, even after analyzing $\sim$10 h of data. Occasionally, failures can be assigned to spectral components not present in the training data. Importantly, the approach requires only moderate data acquisition and training time, and the model can be automatically re-trained if system behaviour changes between days or experimental beamtimes. Overall, we have shown that MLP-driven PID stabilization suggests a way forward for reliable upstream stabilization in such high-energy laser facilities as PHELIX.

The feasibility of deploying this method in a real-time system appears fundamentally sound, provided that several practical aspects are addressed. The required fast steering mirror is standard control hardware and can be procured off the shelf. The only missing component is an experimental characterization of a steering-mirror assembly of the necessary size. Our second-order actuator model, extrapolated from datasheet specifications (mirror mass and drive force), suggests comparable dynamic timescales once such a unit is characterized. The latency budget assumed in this study follows prior AO work at Apollon \cite{Ohland2025}, where high-speed wavefront sensing and a real-time computer (RTC) operated at the relevant bandwidths; a similar platform already exists at PHELIX and can be extended in both software and hardware, with the sole structural addition being the inference latency, currently estimated at \(\sim 0.3\)\ ms, which combined with the \(\sim 0.6\)\ ms system delay remains well within the prediction capabilities of the model. A more fundamental integration challenge arises from the fact that the model assumes direct access to the raw perturbed tip/tilt signal, whereas in most laser systems the diagnostic sensor is located downstream of the steering mirror. This can be addressed either through an open-loop scheme—placing the sensor upstream of the actuator but downstream of the jitter source, supported by an accurate dynamical model of the mirror and optionally a secondary sensor for tuning—or through a pseudo-open-loop architecture that reconstructs the undisturbed input from corrected measurements and historical mirror commands using the actuator model, the latter offering greater generality at the cost of increased control-design complexity. Finally, an experimental evaluation path is available through the ALADIN project (funded by the German BMFRT), which is establishing the LAOTSE beam-control test bench at PHELIX, providing a suitable environment for realistic real-time performance assessment once the required hardware is in place.

%
%

\ack{We gratefully acknowledge support from the ELI-RO/DFG/2023 001 ARNPhot project, funded by the Institute of Atomic Physics (IFA), Romania.

\noindent\textbf{AI assistance statement}

We used OpenAI’s ChatGPT-5 at an early stage for structural feedback and grammar suggestions; subsequent versions have undergone extensive manual editing. The research content, analyses, and conclusions are entirely original and were conceived, validated, and approved by the authors, who take full responsibility for the final manuscript.
}

\funding{This project has received funding by the European Union’s HORIZON-INFRA-2022-TECH-01 call under grant agreement number 101095207. THRILL — Technological and Higher-power laser Research Infrastructures Laboratories
}

\roles{
\textbf{Jiaying Wang:} Conceptualization, Methodology, Software, Formal analysis, Visualization, Writing – original draft, Writing – review \& editing. 

\textbf{Jonas Benjamin Ohland:} Investigation, Data curation, Validation, Writing – review \& editing. 

\textbf{Yen-Yu Chang:} Discussion, Methodology support. 

\textbf{Vedhas Pandit:} Visualization, Writing – review \& editing, Figure preparation, Conceptual discussion. 

\textbf{Stefan Bock:} Methodology support, Supervision (laser), Writing – review \& editing. 

\textbf{Andrew-Hiroaki Okukura:} Investigation, Data curation. 

\textbf{Udo Eisenbarth:} Investigation, Resources, Data curation. 

\textbf{Arie Irman:} Methodology support, Discussion, Writing – review \& editing. 

\textbf{Michael Bussmann:} Supervision, Funding acquisition, Conceptual discussion, Writing – review \& editing. 

\textbf{Ulrich Schramm:} Supervision, Funding acquisition, Conceptual discussion, Writing – review \& editing. 

\textbf{Jeffrey Kelling:} Supervision, Project administration, Methodology, Writing – review \& editing.
}

\data{All datasets and code supporting the findings of this study are openly available in the RODARE repository under the Creative Commons Attribution 4.0 International (CC BY 4.0) license.

The experimental PHELIX pointing data are available at 
\href{https://doi.org/10.14278/rodare.4034}{DOI: 10.14278/rodare.4034}, 
and the corresponding Patch-MLP model files and simulation scripts are archived at 
\href{https://doi.org/10.14278/rodare.4036}{DOI: 10.14278/rodare.4036}.}

The dataset contains 111 one-minute files, each recording the x–y beam center position and corresponding timestamps. The first file is used for training, and the remaining 110 files serve as the test set. The accompanying code release includes a pretrained model, data-loading and visualization scripts, and separate training and testing scripts, all with clear annotations.
\suppdata{
\begin{figure}[H]
    \centering
    \includegraphics[width=0.7\linewidth]{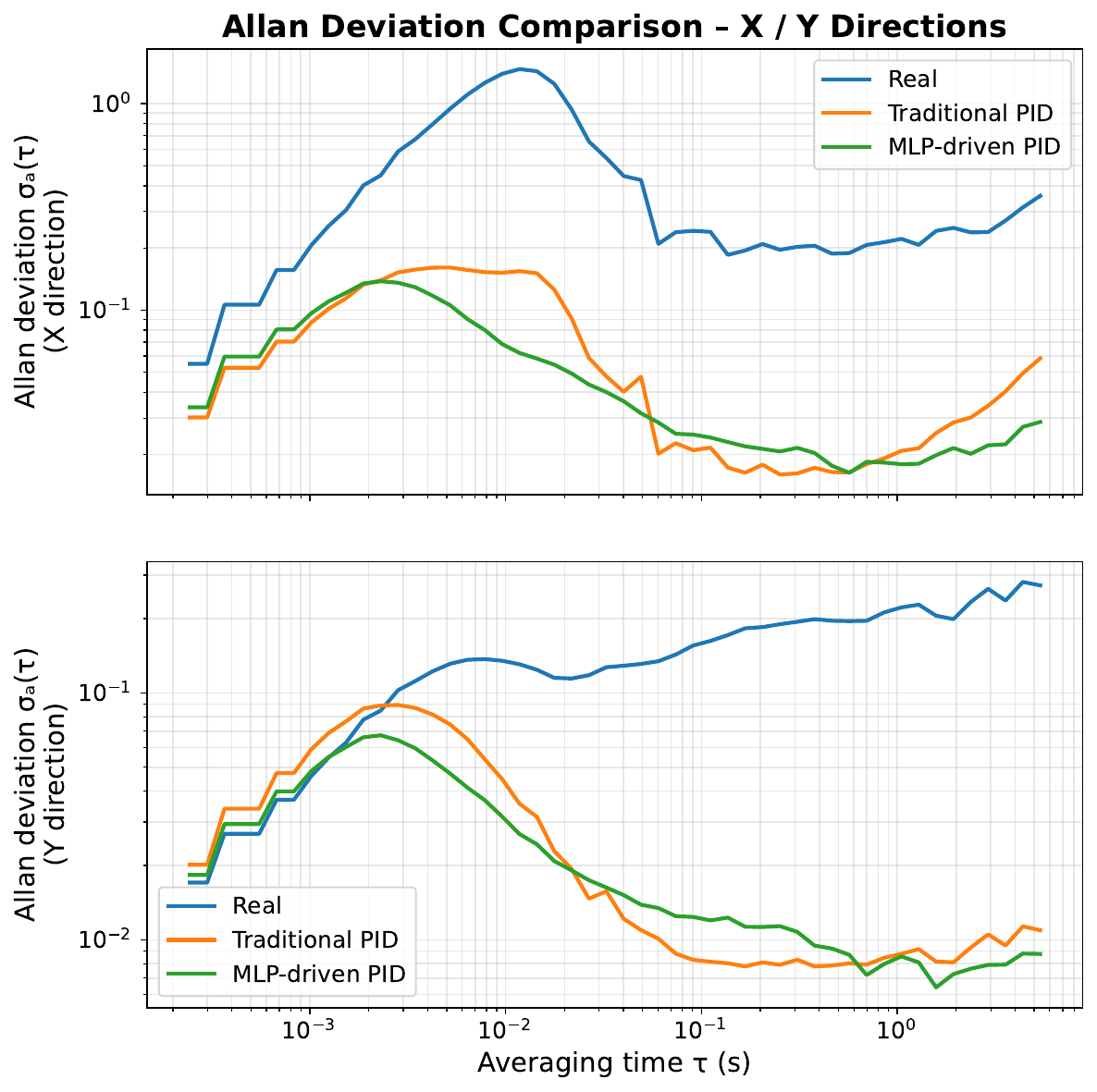}
    \caption{Allan Deviation Comparison: Real Data vs. Simulated PID and Simulated MLP-Driven PID (1 min).}
    \label{fig:allan}
\end{figure}

\begin{figure}[H]
    \centering
    \includegraphics[width=0.7\linewidth]{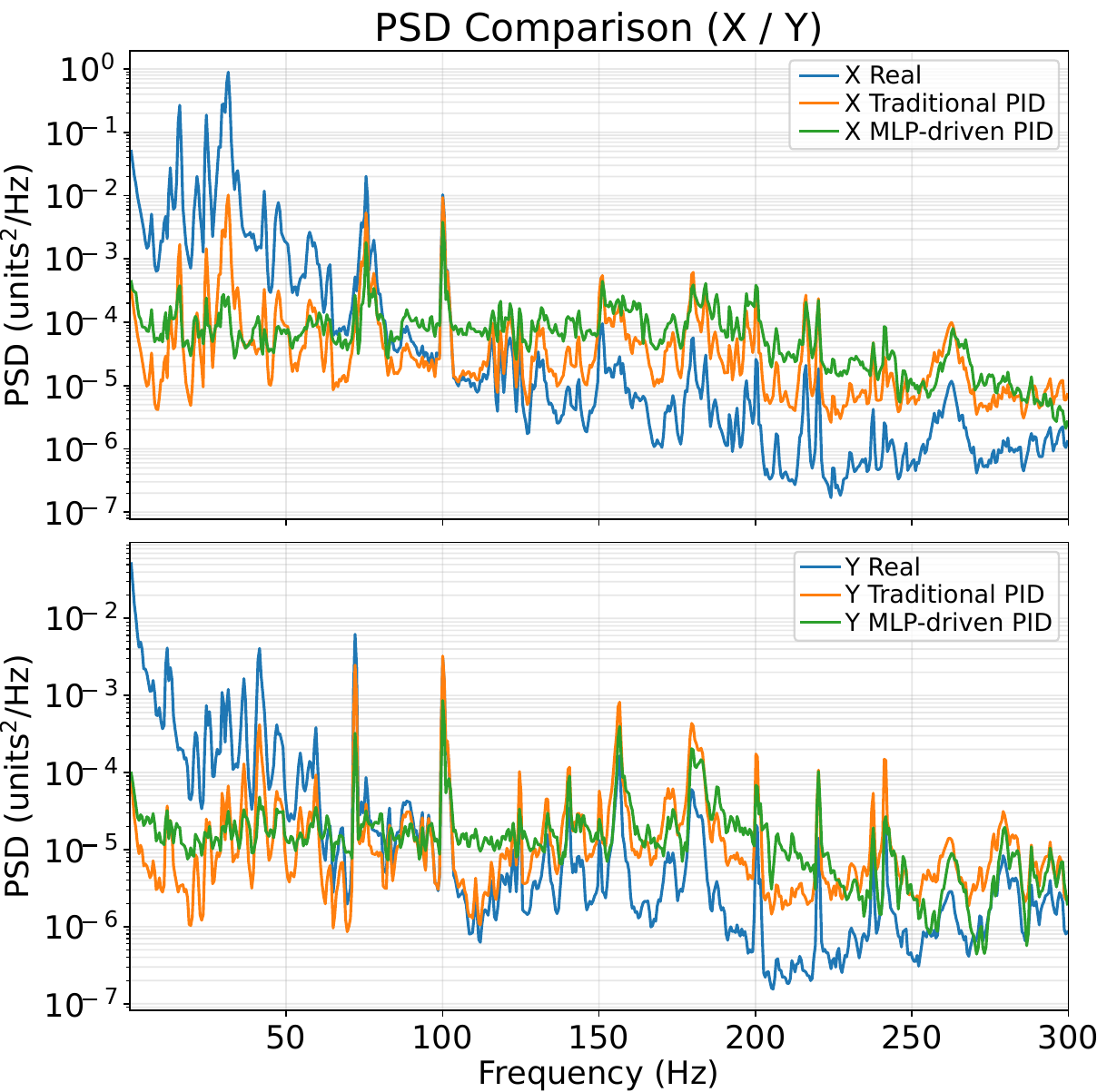}
    \caption{Power spectral density of beam jitter (X/Y, 0–300 Hz).}
    \label{fig:psd_ylog}
\end{figure}

\begin{figure}[H]
    \centering
    \includegraphics[width=0.7\linewidth]{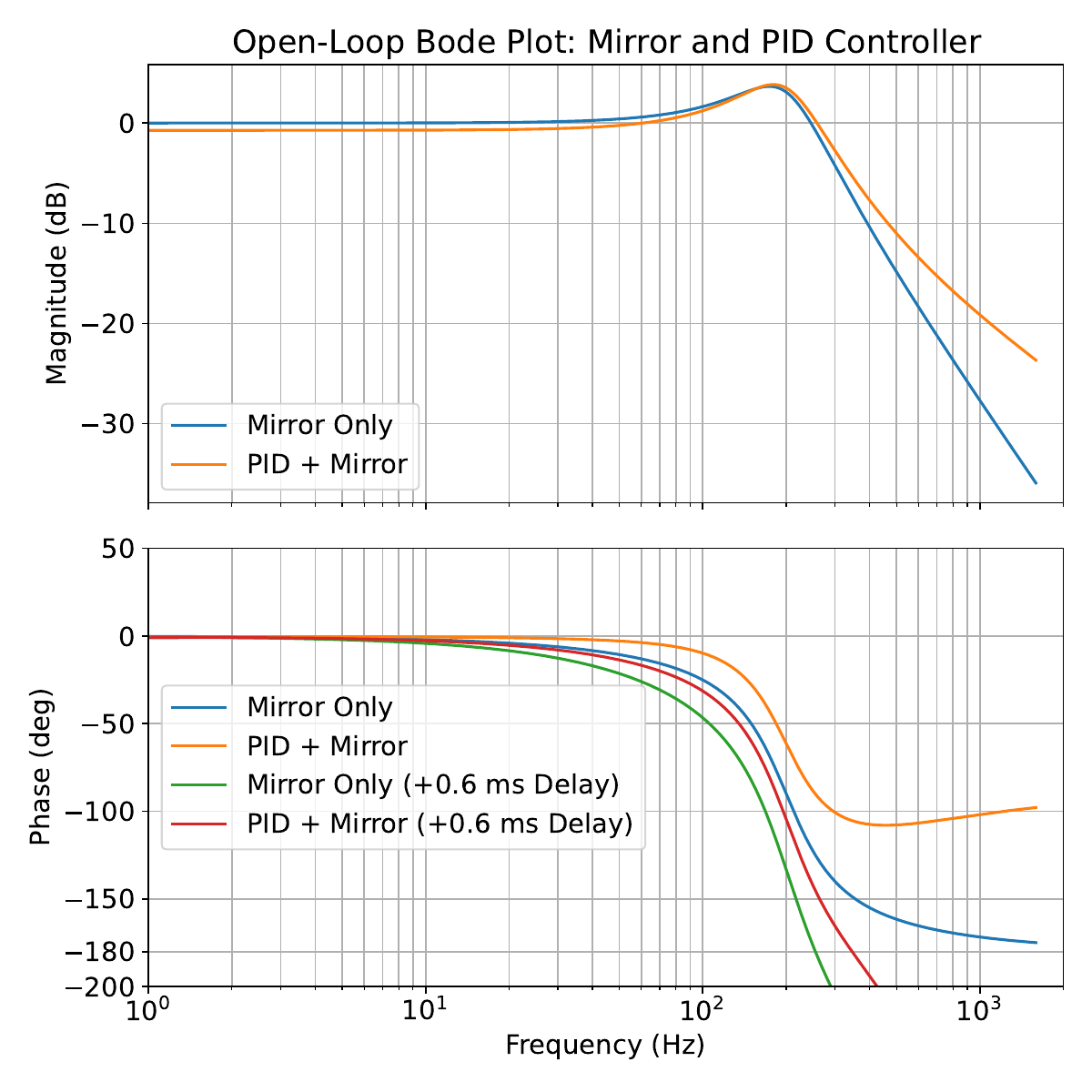}
    \caption{Open-Loop Bode Plot: Mirror and PID Controller.}
    \label{fig:Bode}
\end{figure}

\begin{figure}[H]
    \centering
    \includegraphics[width=1\linewidth]{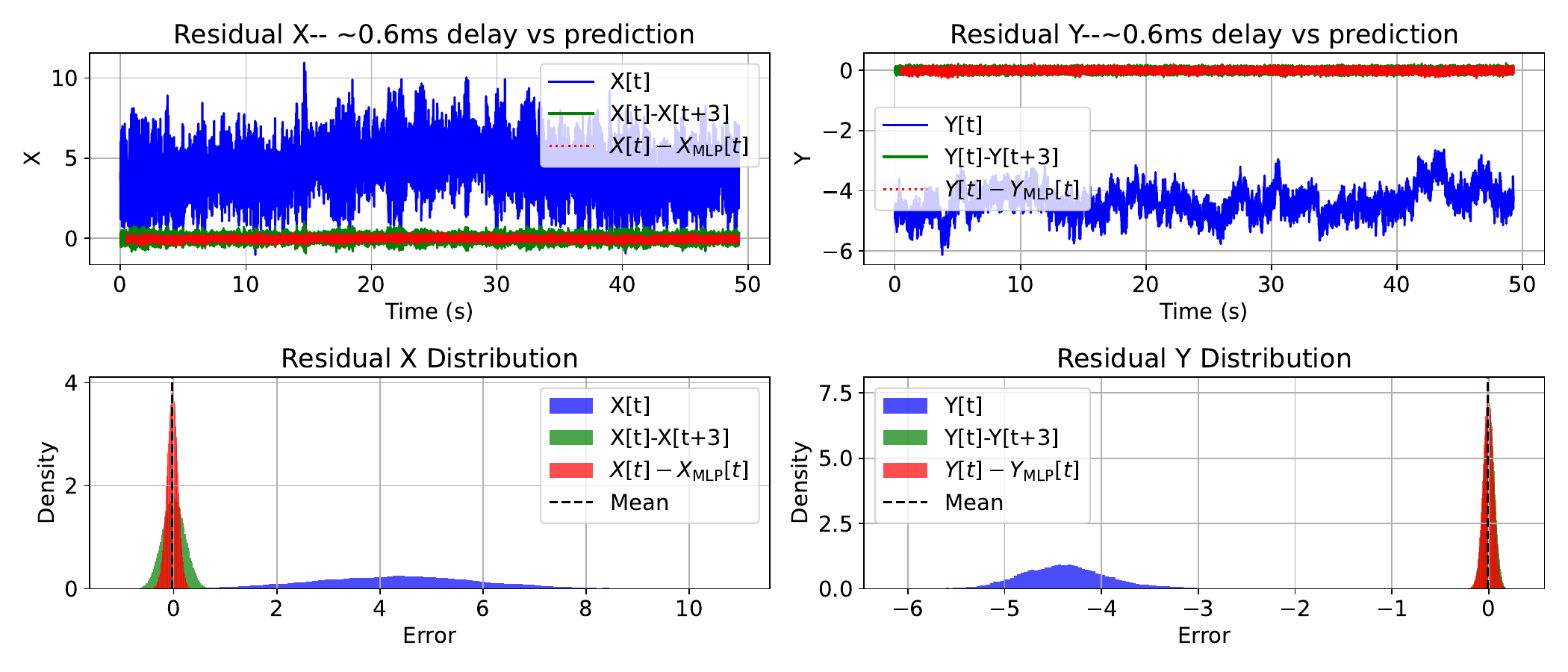}
    \caption{Error in predicted value calculation vs. Error caused by a 0.6ms delay.}
    \label{fig:error}
\end{figure}

\begin{figure}[H]
    \centering
    \includegraphics[width=0.7\linewidth]{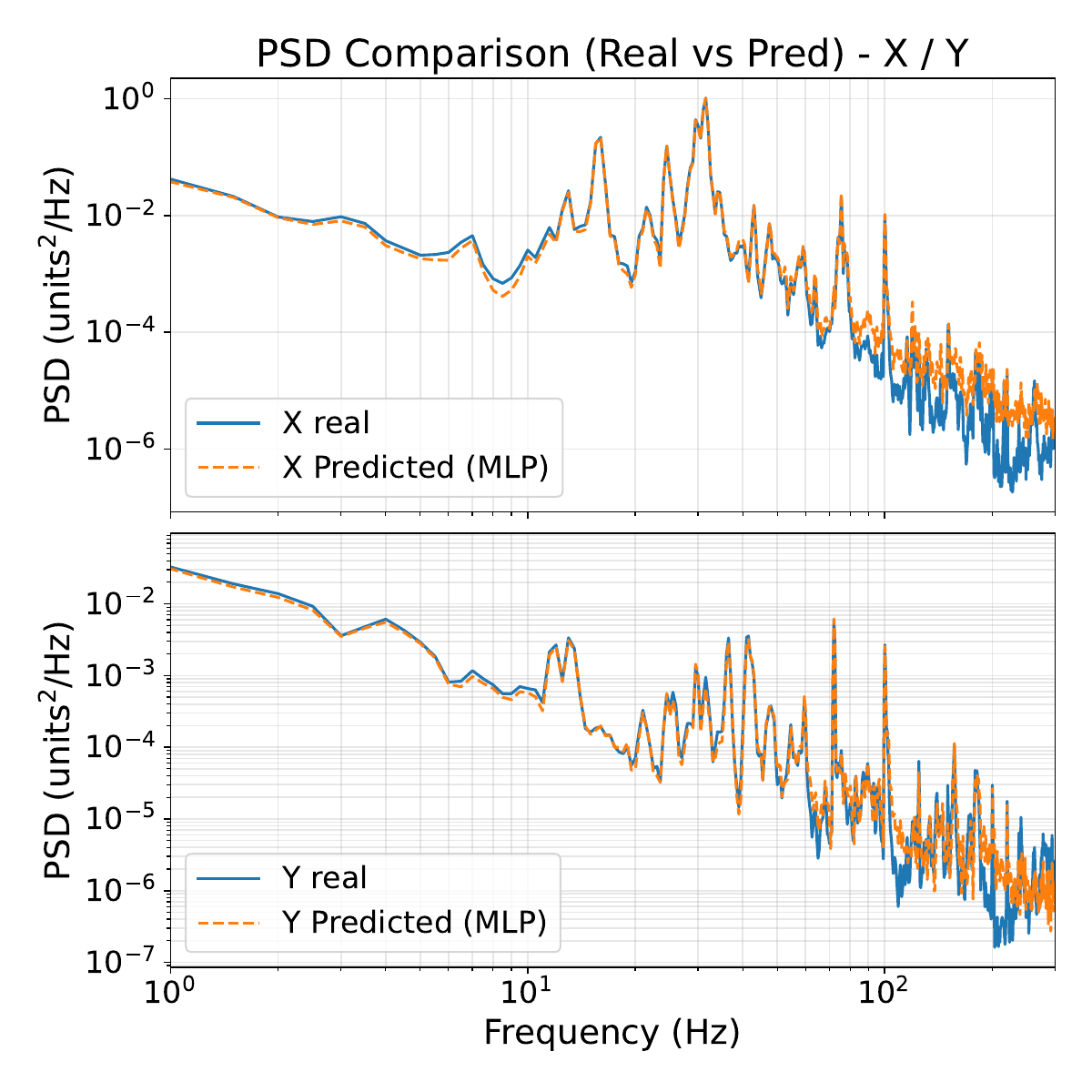}
    \caption{Power spectral density of beam jitter (without simulation).}
    \label{fig:psd without sim}
\end{figure}

\begin{figure}[H]
    \centering
    \includegraphics[width=0.7\linewidth]{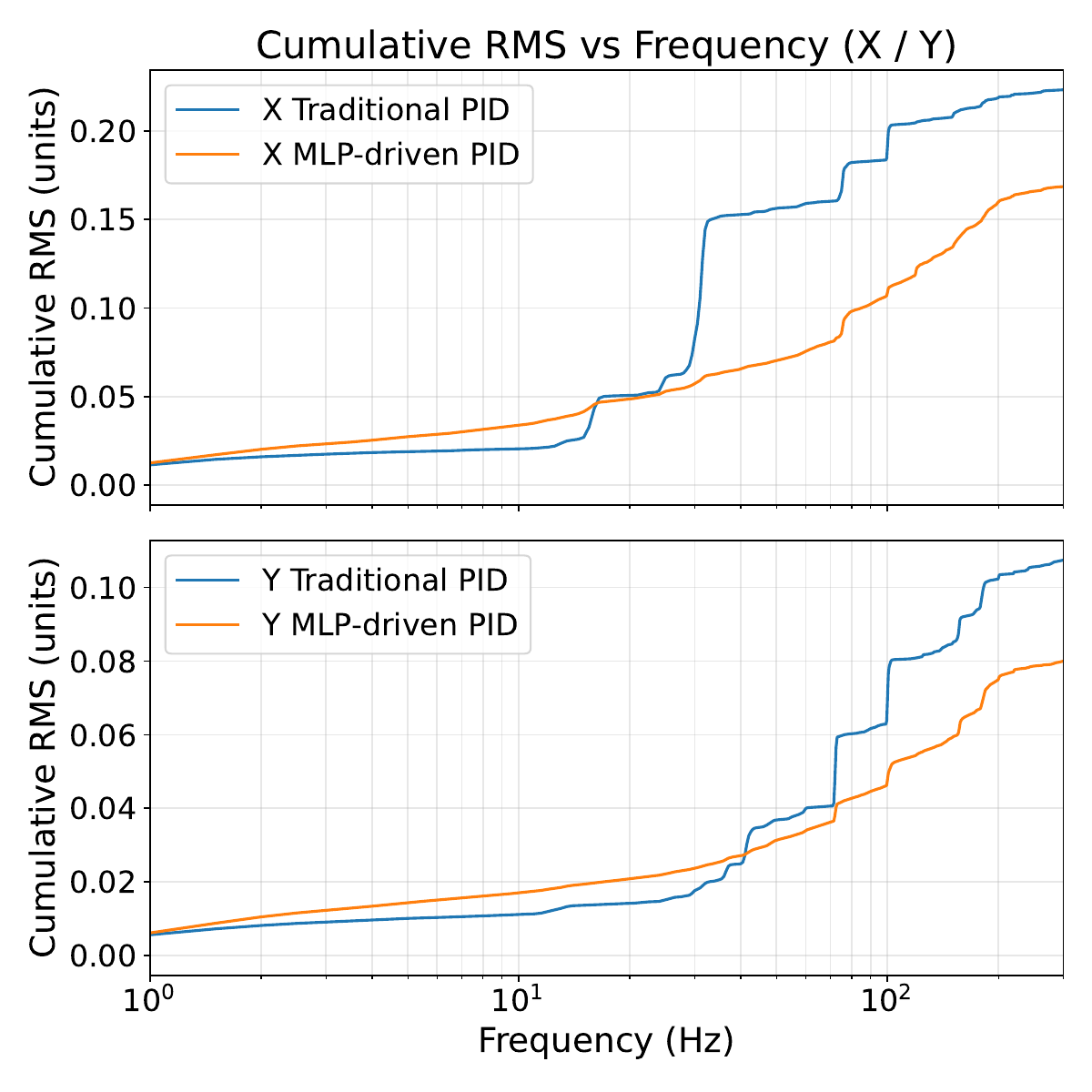}
    \caption{Cumulative RMS of Residual Jitter: Simulated Traditional PID and Simulated MLP-Driven PID.}
    \label{fig:rms}
\end{figure}

}


\bibliographystyle{iopart-num}  
\bibliography{references}

\end{document}